\documentclass[letter,scriptaddress,twocolumn,prl,showkeys]{revtex4}
    \usepackage{graphicx}
	\usepackage{amsmath}%,amssymb} 
	\usepackage{amssymb}
	\usepackage{makeidx}
	\usepackage{amsfonts}
	\usepackage[ansinew]{inputenc}
	\usepackage[usenames,dvipsnames]{pstricks}
	\usepackage{epsfig}
	\usepackage{pst-grad} % For gradients
	\usepackage{pst-plot} % For axes
	\usepackage{color}
	\usepackage[colorlinks,hyperindex]{hyperref}
	\usepackage{lipsum}
	\usepackage{mathrsfs}
	\hypersetup
	{
		colorlinks,%
		citecolor=black,%
		linkcolor=black,%
		urlcolor=black,%
	}

	\newcommand{\ket}[1]{\left| #1 \right\rangle}
	\newcommand{\bra}[1]{\left\langle #1 \right|}

\makeindex

\begin{document}

\title {Wehrl Entropy Based Quantification of Nonclassicality \\
for Single Mode Quantum Optical States}

\author{Soumyakanti Bose}
\email{soumyakanti@bose.res.in}
\affiliation{S. N. Bose National Centre for Basic Sciences \\ Block-JD, Sector-III, Salt Lake, Kolkata 700106 \\ India.\\
}

\date{\today}

\begin{abstract}

Nonclassical states of a quantized light are described in terms of Glauber-Sudarshan $P$ distribution which is not a genuine classical probability distribution. 
Despite several attempts, defining a uniform measure of nonclassicality (NC) for the single mode quantum states of light is yet an open task. 
In our previous work [{Phys. Rev. A {\bf 95}, 012330 (2017)}] we have shown that the existing well-known measures fail to quantify the NC of single mode states that are generated under multiple NC-inducing operations. 
Recently, Ivan {\em et. al.} [{Quantum. Inf. Process. {\bf 11}, 853 (2012)}] have defined a measure of non-Gaussian character of quantum optical states in terms of Wehrl entropy.
Here, we adopt this concept in the context of single mode NC. 
In this paper, we propose a new quantification of NC for the single mode quantum states of light
%, in terms of Wehrl entropy. 
%We quantify the NC of any quantum state 
as the difference between the total Wehrl entropy of the state and the maximum Wehrl entropy arising due to its classical characteristics.
This we achieve by subtracting from its Wehrl entropy, the maximum Wehrl entropy attainable by any classical state that has same randomness as measured in terms of von-Neumann entropy.
We obtain analytic expressions of NC for most of the states, in particular, all pure states and Gaussian mixed states. 
%In the case of Gaussian mixed states, we further show that our proposed quantification is proportional to the nonclassical depth of the state.
However, the evaluation of NC for the non-Gaussian mixed states is subject to extensive numerical computation that lies beyond the scope of the current work. 
We show that, along with the states generated under single NC-inducing operations, also for the broader class of states that are generated under multiple NC-inducing operations, our quantification enumerates the NC consistently.

\end{abstract}


\maketitle

\section*{I. Introduction}
\label{sec_intro}

Quantum states of light exhibit several intriguing features such as photon antibunching, sub-Poissonian distribution, oscillatory number distribution etc. \cite{book_qo_vogel}. 
Optical states revealing such characters play central role in optics \cite{book_qo_appl_perina} as well as in various applications in quantum information theory \cite{book_qo_appl_kok, book_qo_appl_agarwal}. 
The notion of nonclassicality (NC) or quantumness of these optical states is based on the associated phase-space distributions, beyond the scope of classical probability theory \cite{book_qo_schleich}. 
Any quantum state of light $\rho$, can be represented in diagonal coherent state basis as \cite{pdist_glauber, pdist_sudarshan}
\begin{equation}
\rho = \int\frac{d^{2}\alpha}{\pi}~P(\alpha,\alpha^{*})~\ket\alpha\bra\alpha.
\label{eq_def_rho_pdist}
\end{equation}

The state $\rho$ is said to be classical if the Glauber-Sudarshan P distribution behaves like a classical probability distribution, i.e., positive semi-definite or singular no more than a delta function; otherwise nonclassical. 
These nonclassical states could be generated by various NC-inducing operations such as photon excitation \cite{nc_pa_agarwal}, quadrature squeezing \cite{nc_qs}, kerr squeezing \cite{nc_ampsq_yamamoto} etc., applied on the classical states.

Several attempts have been made to quantify the NC of single mode quantum state of light, described in terms of the distance from the nearest classical state in Hilbert space \cite{ncm_trn_hillery, ncm_hsd_dodonov} as well as the sigularity/negativity \cite{ncm_depth_lee, ncm_depth_crit_barnett, ncm_nwf_kenfack} of the associated phase-space distributions. 
People have also proposed interesting approach based on the Monge distance \cite{ncm_md_zyczk}. 
A comprehensive review could be found in \cite{nc_rev_1, nc_rev_2}.
It is well known that single mode nonclassical states are {\em necessary and sufficient} to generate entanglement at the output of a linear device like beam splitter (BS) \cite{nc_bsent_nec, nc_bsent_suff_ivan}. 
This leads to the quantification of input single mode NC in terms of BS output entanglement monotones \cite{ncm_ln_asboth}. 
A relative comparison between Wigner negativity and entanglement potential as measure of single mode NC could be found in \cite{ncm_nwf_ln_compare_li}. 
In the context of conversion of NC into entanglement by BS, for pure states, Vogel and Sperling \cite{nc_bsent_qsp_vogel} have further defined a unified quantification of input NC and output entanglement using the quantum superposition principle.
Recently, people have also suggested quantification of the single mode NC or quantumness in terms of the negativity of the expectation values of normal ordered operators \cite{ncm_noo_gehrke, ncm_noo_nori}.

In our previous work \cite{nc_bsent_bose_kumar}, we have shown that the distance based \cite{ncm_hsd_dodonov} as well as the phase-space based \cite{ncm_depth_lee, ncm_nwf_kenfack} measures fail to capture the NC of single mode quantum states that are generated under multiple NC-inducing operations, reasonably. 
The existing well-known measures \cite{ncm_hsd_dodonov, ncm_depth_lee, ncm_nwf_kenfack} can not account for the relative competition between the NC-inducing operations. 
On the other hand, the operational approach by Gehrke {\em et. al.} \cite{ncm_noo_gehrke}, also finds all the photon number states maximally nonclassical and a squeezed state maximally nonclassical at a moderate squeezing strength. 
Moreover, recent results also show that the single mode NC, defined in terms of entanglement monotones, depends upon the specific choice of entanglement potential \cite{ncm_ep_crit_nori}. This necessitates the search for a {\em consistent measure of NC} for the single mode quantum optical states.

In the current paper, we propose a new quantification of the NC of a single mode quantum optical state in terms of the Wehrl entropy \cite{we_wehrl}. 
Recently, Ivan {\em et. al.} \cite{ngm_we_ivan} have proposed a quantification of non-Gaussian character of any state in terms of Wehrl entropy. 
Here, we adopt this entropic description in the context of NC.
Any state of a quantized electromagnetic field contains both classical and quantum features. 
We quantify the NC of any quantum state of light $\rho$ by subtracting its maximal classical Wherl entropy from its total Wehrl entropy. 
This surplus Wehrl entropy could be interpreted as the Wehrl entropy of $\rho$ arising solely due to its {\em nonclassical/quantum} character. 
As the maximum classical Wehrl entropy of $\rho$ we consider the supremum of the Wherl entropy of all classical states that have the purity equal to $\rho$ itself, as measured by von-Neumann entropy.

We obtain analytic expressions for the NC of both pure states and mixed Gaussian states. 
We further show that our proposed quantification of NC for Gaussian mixed states, is proportional to the nonclassical depth of the state.
However, the evaluation of NC for mixed non-Gaussian states is subject to extensive numerical computation which is not germane to the present paper. 
In the case of quantum pure states, generated under single NC-inducing operation, namely photon number state and quadrature squeezed coherent state, we show that the proposed quantification works fine. 
We also observe that the current quantification describes the NC of states generated by quantum superposition such as Schrodinger cat states, consistently. 
In the case of single mode quantum states, generated under multiple NC-inducing operations, we further obtain NC in line of the concerned BS generated entanglement \cite{nc_bsent_bose_kumar}, {\em in contrast to the existing measures}. 
On the other hand, in the case of mixed Gaussian states also, it is noteworthy that we successfully detect NC only if the state is quadrature squeezed, as reported earlier \cite{ncm_ln_asboth}.

This paper is organized as follows. In {\bf Sec. II}, we introduce the mathematical description of the quantification of NC of single mode quantum optical states, $\mathscr{N}_{\rm{w}}$. 
In {\bf Sec. III}, we derive analytic expressions of $\mathscr{N}_{\rm{w}}$ for pure states and a generic Gaussian state. 
We describe certain properties of $\mathscr{N}_{\rm{w}}$ in {\bf Sec. IV}. 
In {\bf Sec. V} we evaluate the NC for some examples of pure states of a quantized electromagnetic field. 
We discuss the NC of mixed Gaussian states of light in {\bf Sec. VI}. 
Finally, we conclude our work in {\bf Sec. VII}.

\section*{II. Quantifying the Nonclassicality: Defining $\mathscr{N}_{\rm{w}}(\rho)$}

It is well-known that any quantum optical state, $\rho$, contains both classical and quantum characters. 
%We propose a quantification of the nonclassical/quantum character of $\rho$ in terms of its Wehrl entropy, $H_{\rm{w}}(\rho)$ \cite{we_wehrl}. 
We quantify the NC of $\rho$ as the Wehrl entropy that $\rho$ possesses in addition to the Wehrl entropy arising due to its classical character. 
This surplus Wehrl entropy could be interpreted as arising solely due to its quantum/nonclassical character. 
This we achieve by subtracting the maximum Wehrl entropy attainable by a classical state that {\em corresponds} to $\rho$ from the Wehrl entropy of $\rho$ itself. 
The correspondence of the classical state with $\rho$ is discussed next.

%In this section we introduce the quantification of the NC of any single mode quantum optical state, $\mathscr{N}_{\rm{w}}(\rho)$, in terms of the Wehrl entropy of the state.
%Our quantification is based on the surplus Wehrl entropy of the state $\rho$ that it possesses in addition to the Wehrl entropy arising due to its classical characteristics.

\subsection*{II-A. Quantification of Nonclassical Entropy}

It is well-known that the Wherl entropy of any quantum optical state is bounded below by unity \cite{we_min_lieb}, i.e., $H_{\rm{w}}(\rho)\geq 1$. 
The minimum of $H_{\rm{w}}(\rho)$ is attained for a coherent state $\ket z$, the only pure classical state \cite{cls_pure_hillery}. 
As a consequence, while quantifying the nonclassical character of a pure state ($\rho=\ket\psi\bra\psi$) of light with we must restrict ourselves to the set of all classical pure states. 
This is achieved by considering the classical states that have randomness/mixedness same with that of $\ket\psi\bra\psi$. 
There, are plenty of characterizations of the randomness of any quantum state $\rho$. 
Here,we choose von-Neumann entropy, $S(\rho)(=\rm{Tr}[\rho\ln\rho])$, as a measure of randomness. 

In line of the analysis of the pure state, in general, for any nonclassical mixed state $\rho$, we characterize the set of classical reference states by considering only those states which are {\em equientropic} to $\rho$. 
This is what we mean by the {\em correspondence} of the classical states with $\rho$. 
As a consequence, the surplus Wehrl entropy of any state $\rho$ is given by
\begin{equation}
H_{\rm{w}}^{\rm{s}}(\rho)=H_{\rm{w}}(\rho)-\sup_{\sigma\in\Omega_{\rm{cl}}}H_{\rm{w}}(\sigma),
\label{eq_def_we_surplus}
\end{equation}
where, $\Omega_{\rm{cl}}$ is the set of all classical states, s.t. $S(\sigma) = S(\rho)~ \forall~\sigma\in \Omega_{\rm{cl}}$. 
The Wehrl entropy is defined as $H_{\rm{w}}(\rho)=-\int\frac{d^{2}z}{\pi}~Q_{\rho}(z)~ \ln Q_{\rho}(z)$, where, $Q_{\rho}(z)$ ($= \bra z\rho\ket z$) is the Husimi-Kano $Q$ distribution of $\rho$. 
%We denote $H_{\rm{w}}^{\rm{s}}(\rho)$ as the surplus Wehrl entropy that $\rho$ possesses in addition to the Wehrl entropy arising due to its classical character.

\subsection*{II-B. Understanding the Surplus Wehrl entropy ($H_{\rm{w}}^{\rm{s}}$): Quantification of NC ($\mathscr{N}_{\rm{w}}$)}

Let us consider the set of all classical states, $\Omega_{\rm{cl}}:= \lbrace\sigma_{i}\rbrace$ s.t. $S(\sigma_{i}) = S(\sigma_{j})~\forall~\sigma_{i},\sigma_{j}\in\Omega_{\rm{cl}}$. 
Let us further consider two elements from $\Omega_{\rm{cl}}$ as $\sigma_{i}$ and $\sigma_{\rm{max}}$, where
\begin{equation}
\sup_{\sigma\in\Omega_{\rm{cl}}} H_{\rm{w}}(\sigma) = H_{\rm{w}}(\sigma_{\rm{max}}).
\end{equation} 

Now, we replace $\rho$ by $\sigma_{i}$ in Eq. (\ref{eq_def_we_surplus}), i.e., we check for the surplus Wehrl entropy of a classical state. 
This leads to the result
\begin{align}
H_{\rm{w}}^{\rm{s}}(\sigma_{i})&=H_{\rm{w}}(\sigma_{i})-\sup_{\sigma\in\Omega_{\rm{cl}}}H_{\rm{w}}(\sigma) \nonumber \\
&=H_{\rm{w}}(\sigma_{i})-H_{\rm{w}}(\sigma_{\rm{max}})\leq 0.
\end{align}
%\begin{equation}
%H_{\rm{w}}^{\rm{s}}(\sigma_{i})=H_{\rm{w}}(\sigma_{i})-H_{\rm{w}}(\sigma_{\rm{max}})\leq 0.
%\label{eq_we_surplus_cls}
%\end{equation}

%This is not a unexpected result, since, two classical states having same randomness might yield different Wehrl entropy based on their.
Consequently, we define the quantification of NC for any quantum optical state $\rho$ as
\begin{equation}
\mathscr{N}_{\rm{w}}(\rho)=\max\big\lbrace 0,H_{\rm{w}}(\rho)-\sup_{\sigma\in\Omega_{\rm{cl}}}H_{\rm{w}}(\sigma)\big\rbrace.
\label{eq_def_meas_ncwe}
\end{equation}

\section*{III. $\mathscr{N}_{\rm{w}}$ for a Pure and Mixed States of Light}

\subsection*{III-A. $\mathscr{N}_{\rm{w}}$ for Pure States of Light}

It is already discussed that in the case of pure states, the set of reference classical states has to be restricted to the pure states and the only pure classical state is a coherent state \cite{cls_pure_hillery}. 
Moreover, the Wehrl entropy of any quantum state is always greater than or atleast equal to unity, i.e. $H_{\rm{w}}(\rho)\geq 1$; equality holds only for coherent state \cite{we_min_lieb}. 
Thus, for a pure state, $\rho=\ket\psi\bra\psi$, the measure of NC, $\mathscr{N}_{\rm{w}}$, reduces to the analytic form
\begin{equation}
\mathscr{N}_{\rm{w}}(\ket\psi)=H_{\rm{w}}(\ket\psi)~-~1.
\label{eq_meas_ncwe_pure}
\end{equation}

On the other hand, for mixed states, in general, Eq. (\ref{eq_def_meas_ncwe}) has to be respected. 
However, in the case of mixed Gaussian states of light one could obtain analytic expression for the $\mathscr{N}_{\rm{w}}$ by reducing the set of classical reference states further, as discussed next.

\subsection*{III-B. $\mathscr{N}_{\rm{w}}$ for a Gaussian Mixed States of Light}

Any non-Gaussian mixed state of light could be well constructed solely by taking convex combinations of classical states, for example, $\sigma=\frac{1}{2}\big(\ket\alpha\bra\alpha+\ket {-\alpha}\bra {-\alpha}\big)$. 
Such non-Gaussian states could yield arbitrarily high Wehrl entropy based upon the very combination. 
On the other hand, for a Gaussian state, $\rho^{\rm{G}}$, its Wehrl entropy is well defined in terms of its variance matrix (Appendix). 
Thus, in the context of Wehrl entropy of any Gaussian state arising due to its classicality, the classical character of the $\rho^{\rm{G}}$ is best represented by only a Gaussian state. 
As a consequence, in the case of nonclassical Gaussian mixed states, we further restrict our set of classical states in Eq. (\ref{eq_def_meas_ncwe}) to the Gaussian states. 
With this choice of classical reference states, in the case of any Gaussian mixed state $\rho^{\rm{G}}$, the quantity $\mathscr{N}_{\rm{w}}$ in Eq. (\ref{eq_def_meas_ncwe}) reduces to
\begin{equation}
\mathscr{N}_{\rm{w}}(\rho^{\rm{G}})=\max\big\lbrace 0,H_{\rm{w}}(\rho^{\rm{G}})-\sup_{\sigma^{\rm{G}}\in\Omega_{\rm{cl}}}H_{\rm{w}}(\sigma^{\rm{G}})\big\rbrace,
\label{eq_def_meas_ncwe_gauss}
\end{equation}
where "G" stands for Gaussian. 
Next we derive an analytic expression for $\mathscr{N}_{\rm{w}}(\rho^{\rm{G}})$.

Let us consider a single mode Gaussian state of light, $\rho^{\rm{G}}$, described by the variance matrix $V$. 
Corresponding Wigner function is given by,
\begin{equation}
W^{\rm{G}}(R)=\sqrt{\det V^{-1}}~e^{-\frac{1}{2}(R-D)^{\rm{T}}V^{-1}(R-D)},
\label{eq_def_wig_gauss}
\end{equation}
where, $R~\equiv~(x,p)^{\rm{T}}$ is the column vector formed with the real quadrature $(x,p)$. 
The displacement (column) vector is given as $D~\equiv~ (\langle x\rangle,\langle p\rangle)^{\rm{T}}$.
T stands for transposition.
The real symmetric variance matrix $V$ is defined as $V_{\mu,\nu}=\frac{1}{2}\rm{Tr}~[~\lbrace \hat{R}_{\mu},\hat{R}_{\nu}\rbrace~ \rho]~;~\mu,\nu=1,2$.
The operator analogue of $R_{\mu}$ ($R_{1}=x,R_{2}=p$) is $\hat{R}_{\mu}$.
The quantum variance matrix $V$ satisfies the canonical uncertainty relation \cite{varmat_simon}
\begin{equation}
V + \frac{i}{2}~\Omega\geq 0,
\label{eq_vm_cond_bonafide}
\end{equation}
where, $\Omega$ is the symplectic metric given as
defined as $\Omega=\begin{pmatrix}
0 & 1\\
-1 & 0
\end{pmatrix}$.

By using the transformation relation between the $Q$ distribution and Wigner distribution ($W$) it can be shown by straightforward calculation that the Wehrl entropy of the Gaussian state $\rho^{\rm{G}}$ is given by (Appendix)
\begin{equation}
H_{\rm{w}}(\rho^{\rm{G}})=1+\frac{1}{2}\ln[\frac{4\delta+2\epsilon+1}{2}],
\label{eq_we_gauss}
\end{equation}
where, $\delta=\det V$ and $\epsilon=\rm{Tr}[V]$.

It is well-known that any single mode Gaussian state of light could be written as a displaced squeezed thermal state \cite{smgs_dst_chaturvedi}
\begin{equation}
\rho^{\rm{G}}~=~D(\alpha)S_{\rm{sq}}(z)\rho_{\rm{th}}(\bar{n})S^{\dagger}_{\rm{sq}}(z)D^{\dagger}(\alpha),
\label{eq_def_smgs_dst}
\end{equation}
where, $S_{\rm{sq}}(z)=e^{\frac{1}{2}(z a^{\dagger 2}-z^{*}a^{2})}$, $z~=re^{i\theta}$ and $D(\alpha)=e^{\alpha a^{\dagger}-\alpha^{*}a}$.
In the expression of squeezing operator, $S_{\rm{sq}}(z)$, $r$ defines the degree of squeezing and $\theta$ is the angle of squeezing.
The von-Neumann entropy of the Gaussian state in Eq. (\ref{eq_def_smgs_dst}) is given as \cite{varmat_simon, entropy_gauss_illuminati}
\begin{equation}
S(\rho^{\rm{G}})=S(\rho_{\rm{th}}(\bar{n}))=(\bar{n}+1)\ln(\bar{n}+1)-\bar{n}\ln\bar{n}.
\label{eq_gauss_entropy}
\end{equation}

The symplectic eigenvalue of the variance matrix $V$ (in the Williamson's diagonal form) for the single mode Gaussian state $\rho^{\rm{G}}$ is related to the average thermal photon as $\lambda~=~\bar{n}~+~\frac{1}{2}$.
This symplectic eigenvalue could be obtained as the ordinary eigenvalue of the matrix $|i V\Omega|$.

Now the question arises is about choosing the classical Gaussian reference, with the same von-Neumann entropy, that yields maximum Wehrl entropy. 
Similar to the description given above, we can choose the thermal state in the classical reference state by equating its entropy with that of the Gaussian state $\rho^{\rm{G}}$. 
Furthermore, any single mode Gaussian state of the form given in Eq. (\ref{eq_def_smgs_dst}) is classical for $r\leq~\frac{1}{2}~\ln~(2\bar{n}~+~1)$. 
Hence, the maximum Wehrl entropy attainable by by any classical Gaussian state with entropy equal to $S(\rho^{\rm{G}})$ is achieved for
\begin{equation}
\sigma^{\rm{G}}_{\max}~=~S_{\rm{sq}}(r_{\max})\rho_{\rm{th}}(\bar{n})S^{\dagger}_{\rm{sq}}(r_{\max}),
\label{eq_def_gauss_cls_ref}
\end{equation}
where, $(\bar{n}+1)\ln(\bar{n}+1)-\bar{n}\ln\bar{n}=S(\rho^{\rm{G}})$ and $r_{\max}=\frac{1}{2}\ln(2\bar{n}+1)$.

Using the relation between the classical and quantum state parameters, as described in Eq. (\ref{eq_def_gauss_cls_ref}), it could be easily shown that the quantum variance matrix for the above mentioned classical Gaussian state $\sigma^{\rm{G}}_{\max}$ could be represented in terms of the variance matrix of the Gaussian state $\rho^{\rm{G}}$ as
\begin{equation}
V^{\rm{G}}_{\rm{cl}}~=~\sqrt{\delta}\begin{pmatrix}
\sqrt{4\delta} & 0\\
0 & 1/\sqrt{4\delta}
\end{pmatrix},
\label{eq_vm_cls_ref}
\end{equation}
where, $\det V^{\rm{G}}_{\rm{cl}}=\det V=\delta$ and $\rm{Tr}[V^{\rm{G}}_{\rm{cl}}]=(4\delta+1)/2$.
The Wehrl entropy for the classical Gaussian reference state described by the variance matrix $V^{\rm{G}}_{\rm{cl}}$ is given by (Eq. \ref{eq_we_gauss})
\begin{equation}
H_{\rm{w}}(\sigma^{\rm{G}}_{\max})=1+\frac{1}{2}\ln[4\delta+1].
\label{eq_we_gauss_cls}
\end{equation}

Consequently, Eq. (\ref{eq_def_meas_ncwe}), (\ref{eq_we_gauss}), (\ref{eq_def_gauss_cls_ref}) and (\ref{eq_we_gauss_cls}) lead to the analytic expression of quantumness of the Gaussian state $\rho^{\rm{G}}$ as
\begin{equation}
\mathscr{N}_{\rm{w}}(\rho^{\rm{G}})=\max\big\lbrace 0,\frac{1}{2}\ln[\frac{4\delta+2\epsilon+1}{2(4\delta+1)}]\big\rbrace.
\label{eq_meas_ncwe_gauss}
\end{equation}

The state $\rho^{\rm{G}}$ is said to nonclassical if $4\delta+2\epsilon+1>2(4\delta+1)$ or $2\epsilon> 4\delta+1$.

\subsubsection*{\bf Relation with Nonclassical Depth}

In this connection it is worth comparing the result of Eq. (\ref{eq_meas_ncwe_gauss}) with nonclassical depth \cite{ncm_depth_lee} that serves a good measure of NC for single mode Gaussian states \cite{ncm_depth_crit_barnett}. 
A single mode Gaussian state $\rho^{\rm{G}}$ is said to be nonclassical if it has a non-zero depth. 
The nonclassical depth for $\rho^{\rm{G}}$ is given by
\begin{equation}
\eta~=\max\big\lbrace 0, \frac{1}{2}-\lambda_{\min}\big\rbrace,
\label{eq_def_depth_gauss}
\end{equation}
where, $\lambda_{\min}$ is the minimum eigenvalue of the variance matrix $V$.
It is given by
\begin{equation}
\lambda_{\min}=\frac{\epsilon-\sqrt{\epsilon^{2}-4\delta}}{2}.
\label{eq_def_lmin_vm_gauss}
\end{equation}

AS evident from Eq. (\ref{eq_def_depth_gauss}) and (\ref{eq_def_lmin_vm_gauss}), the condition of NC of $\rho^{\rm{G}}$, $\eta>0$, yields
\begin{align}
&\lambda_{\min}<\frac{1}{2} \nonumber \\
\Rightarrow&~ 2\epsilon> 4\delta+1  
\label{eq_cond_nc_depth}
\end{align}

This is exactly the condition of NC for $\rho^{\rm{G}}$ that one derives from Eq. (\ref{eq_meas_ncwe_gauss}).

\subsection*{III-B. $\mathscr{N}_{\rm{w}}$ for a non-Gaussian Mixed States of Light}

In the case of non-Gaussian mixed states, however, one needs to consider the set of all {\em equientropic} classical states as mentioned in the Eq. (\ref{eq_def_meas_ncwe}).
In such cases, one has to find the supremum over all possible states which is subject to heavy numerical computation and lies beyond the scope of the current work. 
This, we shall consider elsewhere.
In the current work, we focus on the analytic evaluation of the NC of single mode quantum optical states with the proposed quantification based on Wehrl entropy.
 
Next, we discuss some properties of the proposed quantification of single mode NC, $\mathscr{N}_{\rm{w}}$, that we shall be using often while evaluating it for different states.

\section*{IV. Some Properties of $\mathscr{N}_{\rm{w}}$}

\subsection*{IV-A. Invariance under Displacement}

Let us consider a any quantum optical state $\rho$, for which the Husimi $Q$ distribution is given as $Q_{\rho}(\beta)=\bra\beta\rho\ket\beta$. 
Under the action of a phase-space displacement, $D(z):\rho\rightarrow\tilde{\rho}= D(z)\rho D^{\dagger}(z)$, Husimi $Q$ distribution changes as
\begin{equation}
D(z):Q_{\rho}(\beta)\rightarrow Q_{\tilde{\rho}}(\beta)= Q_{\rho}(\beta-z).
\label{eq_qdist_invar_displace}
\end{equation}

This indicates that the phase-space displacement works as the rigid translation \cite{we_wehrl, we_min_lieb} that leaves Wehrl entropy unchanged, i.e. $D(z): H_{\rm{w}}(\rho)\rightarrow H_{\rm{w}}(\rho)$. 
Since, the Wehrl entropy of any state is independent of phase-space displacement, it is evident from the Eq. (\ref{eq_def_meas_ncwe}), that under the transformation $D(z): \rho\rightarrow\tilde{\rho}$, its NC does not change, i.e.,
\begin{equation}
D(z):\mathscr{N}_{\rm{w}}(\rho)\rightarrow \mathscr{N}_{\rm{w}}(\rho).
\label{eq_ncwe_invar_displace}
\end{equation}

\subsection*{IV-B. Invariance under Passive Rotation}

Let us now consider a passive rotation in phase space $T_{U}:\rho\rightarrow\tilde{\rho}$, where $U$ is the transformation indicating rotation in the phase-space quadrature. 
Under the transformation $T_{U}$, Husimi $Q$ distribution changes as
\begin{equation}
T_{U}:Q_{\rho}(\beta)\rightarrow Q_{\tilde{\rho}}(\beta)=Q_{\rho}(U^{-1}\beta).
\end{equation}

Since, the Jacobean of the passive rotation in phase-space is unity, we have $T_{U}:H_{\rm{w}}(\rho)\rightarrow H_{\rm{w}}$, i.e., the Wehrl entropy does not change under passive rotation in phase-space. 
This leads to the fact that under a passive phase-space rotation $T_{U}:\rho\rightarrow\tilde{\rho}$, the NC, $\mathscr{N}_{\rm{w}}(\rho)$, as defined in Eq. (\ref{eq_def_meas_ncwe}), does not change.

Here, it is worth mentioning that only in two-dimension $SO(2,R)\subseteq Sp(2,R)$.
That means only in the case of single mode, all proper rotations are canonical transformations. 
However, in the case of multimode, this is not true. 
For a system of $N$ ($N\geq 2$) harmonic oscillators, rotations belong to $SO(2N,R)$ which is a much bigger group than the $Sp(2N,R)$ that characterizes symplectic or canonical transformations.
As a consequence, in multimode, all phase space rotations are not canonical.
In such cases only those rotations which belong to $Sp(2N,R)\cap SO(2N,R)$ are allowed.

Next, we evaluate $\mathscr{N}_{\rm{w}}$ for certain well-known examples of single mode nonclassical pure as well as Gaussian mixed states.

\section*{V. $\mathscr{N}_{\rm{w}}$ for Some Pure States}

\subsection*{V-A. Photon Number State and Quadrature Squeezed Coherent State:}

A photon number state $\ket m$ is obtained by applying photon excitation $\big( \frac{a^{\dagger m}}{\sqrt{m!}}\big)$ on the vacuum.
The Wehrl entropy of a photon number state For $\ket m$ is given by \cite{ngm_we_ivan}
\begin{equation}
H_{\rm{w}}(\ket m)=1+m+\ln m!-m\Psi(m+1),
\label{eq_we_fock}
\end{equation}
where, $\Psi(m+1)=\sum_{k=1}^{m}\frac{1}{k}-\gamma$ is the di-gamma function.
The Euler constant $\gamma$ is given as $\gamma = 0.5722..$.
This leads to the analytic expression for NC of $\ket m$ as
\begin{equation}
\mathscr{N}_{\rm{w}}(\ket m)=m+\ln m!-m\Psi(m+1).
\label{eq_meas_ncwe_fock}
\end{equation}

\begin{figure}[h!]
\includegraphics[scale=1.2]{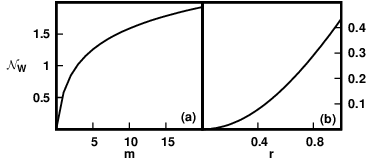}
\caption{Plot of $\mathscr{N}_{\rm{w}}$ for {\bf (a)} Photon number state and {\bf (b)} Squeezed coherent state.}
\label{fig_ncwe_fsv}
\end{figure}

We plot the $\mathscr{N}_{\rm{w}}(\ket m)$ for different values of $m$ in Fig. \ref{fig_ncwe_fsv}(a). 
It increases monotonically with the increase in number of photon addition $m$.
For small $m(\leq 4)$ we observe a rapid increase in $\mathscr{N}_{\rm{w}}(\ket m)$.
With further increase in $m$, $\mathscr{N}_{\rm{w}}(\ket m)$ saturates for very high $m$. 
It is noteworthy that the monotonic increase in $\mathscr{N}_{\rm{w}}(\ket m)$ with increase in $m$ falls in line of the increasing negativity in the Wigner distribution \cite{ncm_nwf_kenfack}.

A quadrature squeezed coherent state, $\ket {\psi_{\rm{sc}}}=S(\zeta)\ket\alpha$, is generated under quadrature squeezing, $S(\zeta)=\exp\big( \frac{\zeta a^{\dagger 2}~-~\zeta^{*}a^{2}}{2}\big)$, applied on a coherent state $\ket\alpha$, where $\zeta=r e^{i\theta}$; $r$ and $\theta$ being the squeezing strength and the squeezing angle respectively. 
For $\ket {\psi_{\rm{sc}}}$ we obtain a logarithmic NC as
\begin{equation}
\mathscr{N}_{\rm{w}}(\ket {\psi_{\rm{sc}}})=\ln \mu,
\label{eq_meas_ncwe_sc}
\end{equation}
where $\mu = \cosh r$. 
Evidently, the $\mathscr{N}_{\rm{w}}(\ket {\psi_{\rm{sc}}})$ is independent of $\theta$, since it only sets the direction of squeezing rather than the degree of squeezing. 
Moreover, the $\mathscr{N}_{\rm{w}}(\ket {\psi_{\rm{sc}}})$ is independent of the coherent displacement $\alpha$, e.g., $\mathscr{N}_{\rm{w}}(S(\zeta)\alpha)=\mathscr{N}_{\rm{w}}(S(\zeta)\ket 0)$. 
This can be explained in the following way. 
The state $\ket {\psi_{\rm{sc}}}$ could be written as
\begin{equation}
\ket {\psi_{\rm{sc}}}= S(\zeta)\ket\alpha= S(\zeta)D(\alpha)\ket 0= D(\beta)S(\zeta)\ket 0,
\end{equation} 
where, $\beta=\mu\alpha-\nu e^{i\theta}\alpha^{*}$, $\mu=\cosh r$ and $\nu=\sinh r$.
We have already discussed that the Wehrl entropy is independent of displacement in phase space.
As a consequence, the NC of $\ket {\psi_{\rm{sc}}}$ is independent of the coherent displacement $\alpha$. 
In Fig. \ref{fig_ncwe_fsv}(b) we plot the dependence of $\mathscr{N}_{\rm{w}}(\ket {\psi_{\rm{sc}}})$ upon $r$. 
We observe an initial slow and then rapid increase in $\mathscr{N}_{\rm{w}}(\ket {\psi_{\rm{sc}}})$ with increase in $r$. 
However, for very high $r$ it saturates asymptotically (not shown in the figure).

\subsection*{V-B. Photon Added Coherent State and Coherent Superposition States:}

\begin{figure}[h!]
\hspace*{-10pt}
\includegraphics[scale=1.7]{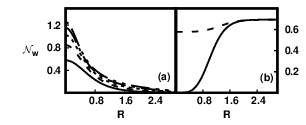}
%\vspace*{-10pt}
\caption{Plot of dependence of $\mathscr{N}_{\rm{w}}$ on $R$ for {\bf (a)} $\ket{\psi_{\rm{pac}}}$ for $m=1$ (solid line), $2$ (dashed line), $3$ (dotted line), $4$ (dashed dotted line) and $5$ (dashed double dotted line) {\bf (b)} $\ket{\psi_{\pm}}$ with $\ket{\psi_{+}}$ (solid line) and $\ket{\psi_{-}}$ (dashed line).}
\label{fig_ncwe_pac_cs}
\end{figure}

An m-photon added coherent state (PAC) is given as
\begin{equation}
\ket{\psi_{\rm{pac}}}=\frac{1}{\sqrt{C_{m}}}~a^{\dagger m}~\ket\alpha,
\label{eq_psi_pac}
\end{equation}
where, $C_{m}=m!~L_{m}(-|\alpha|^{2})$ is the normalization constant.   
For the sake of simplicity we consider real displacement, i.e., $\alpha=R$. 
In Fig. \ref{fig_ncwe_pac_cs}(a) we plot the dependence of $\mathscr{N}_{\rm{w}}\big(\ket{\psi_{\rm{pac}}}\big)$ on $R$ for different $m$ values. 
With increase in $m$, $\mathscr{N}_{\rm{w}}\big(\ket{\psi_{\rm{pac}}}\big)$ increases monotonically that signifies the increasing NC.
On the other hand as $R$ increases $\mathscr{N}_{\rm{w}}\big(\ket{\psi_{\rm{pac}}}\big)$ decreases monotonically revealing the increase in the classical character of the state. 
For sufficiently high $R$ ($>>1$), $\mathscr{N}_{\rm{w}}\big(\ket{\psi_{\rm{pac}}}\big)$ becomes almost independent of $m$.
This is quite expected, since, for sufficiently large coherent amplitude  

We further study the even ($\ket{\psi_{+}}$) an the odd ($\ket{\psi_{-}}$) superposition of coherent states. 
These states are given as
\begin{equation}
\ket{\psi_{\pm}}=\frac{\ket\alpha \pm \ket{-\alpha}}{\sqrt{2~\big(1 \pm e^{-2|\alpha|^{2}} \big)}}.
\label{eq_psi_cs}
\end{equation}

For the sake of simplicity we consider real displacement, i.e., $\alpha=R$.
We show the dependence of $\mathscr{N}_{\rm{w}}$ on $R$ for $\ket{\psi_{\pm}}$ in Fig. \ref{fig_ncwe_pac_cs}(b). 
It is noteworthy that for small $R$ ($\lesssim 1.0$), $\ket{\psi_{-}}$ is
more nonclassical than $\ket{\psi_{+}}$; however, for large $R$ ($\gtrsim 1.5$), both $\ket{\psi_{\pm}}$ are equally nonclassical. 
This can be explained in the following way.
The Husimi-Kano $Q$ distributions for the even and odd superposition states are given as 
\begin{equation}
Q_{\ket{\psi_{\pm}}}(\beta)=\frac{e^{-R^{2}}e^{-|\beta|^{2}}}{1+2e^{-2R^{2}}}\big( \cosh [2R\beta_{\rm{re}}]\pm \cos [2R\beta_{\rm{im}}] \big),
\label{eq_qdist_cs}
\end{equation}
where, $\beta_{\rm{re}}$ and $\beta_{\rm{im}}$ are the real and complex part of the quadrature variable $\beta$.
In the expression of $Q$ distribution for the $\ket{\psi_{\pm}}$, the second term in the bracket is a circular function that is bounded by $\pm 1$ while the first term is unbounded.
As a consequence, in the large $R$ limit only the first term predominates while the contribution from the second term becomes negligible.
That is to say that in the limit $R\rightarrow \infty$, the $Q$ distributions for both $\ket{\psi_{\pm}}$ in Eq. (\ref{eq_qdist_cs}) reduce to 
\begin{equation}
\lim_{R\rightarrow\infty} Q_{\ket{\psi_{\pm}}}(\beta) \rightarrow \frac{e^{-R^{2}}}{1+2e^{-2R^{2}}}~e^{-|\beta|^{2}}~\cosh [2R\beta_{\rm{re}}].
\label{eq_qdist_cs_large_R}
\end{equation}

As a consequence of Eq. (\ref{eq_qdist_cs_large_R}), with increase in $R$, for both $\ket{\psi_{\pm}}$ we obtain equal NC.

\subsection*{V-C. Photon Added Squeezed Vacuum State and Squeezed Number State:}

\begin{figure}[h!]
%\hspace*{-10pt}
\includegraphics[scale=1.3]{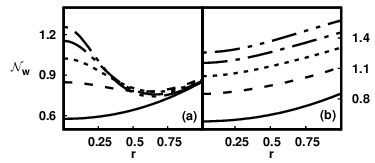}
%\vspace*{-10pt}
\caption{Plot of $\mathscr{N}_{\rm{w}}$ vs $r$ for $m=1$ (solid line), $2$ (dashed line), $3$ (dotted line), $4$ (dashed dotted line) and $5$ (dashed double dotted line) for {\bf (a)} PAS and {\bf (b)} SNS.}
\label{fig_ncwe_pas_sns}
\end{figure}

We have also considered the single mode quantum optical states generated under successive application of multiple NC-inducing operations, in particular, photon excitation and quadrature squeezing. 
The ordered application of these operations on vacuum lead to the states known as photon added squeezed vacuum state (PAS) and squeezed number state (SNS). 
These are given as
\begin{align}
\ket{\psi_{\rm{pas}}}&= \frac{1}{\sqrt{N_{m}}}~a^{\dagger m} S(r) \ket 0 \nonumber \\
\ket{\psi_{\rm{sns}}}&= S(r) \ket m = S(r) \frac{a^{\dagger m}}{\sqrt{m!}} \ket 0,
\label{eq_psi_pas_sns}
\end{align}
where, $N_{m}= m! \mu^{m} P_{m}(\mu)$, $\mu=\cosh r$ and $P_{n}(x)$ is the $n^{\rm{th}}$ order Legendre Polynomial.
In Fig. \ref{fig_ncwe_pas_sns} we have plotted the dependence of $\mathscr{N}_{\rm{w}}$ on the squeeze parameter $r$ for PAS and SNS for different values of $m$.

In the case of PAS [Fig. \ref{fig_ncwe_pas_sns}(a)], we observe that $\mathscr{N}_{\rm{w}}$ is non-monotonic on both $r$ and $m$. 
For $m=1$ it increases monotonically with $r$. 
However, $\forall~ m\geq 2$, as $r$ increases $\mathscr{N}_{\rm{w}}\big(\ket{\psi_{\rm{pas}}} \big)$ first decreases and then increases. 
In the region where the competition between photon excitation and quadrature squeezing becomes explicit ($0.30\leq r\leq 0.60$), $\mathscr{N}_{\rm{w}}\big(\ket{\psi_{\rm{pas}}} \big)$ for higher $m$ becomes smaller than the lower $m$, as expected from the corresponding BS generated entanglement \cite{nc_bsent_bose_kumar}. 
It becomes prominent with increase in $m$. 
For very high value of $r$ ($\gtrsim 0.80$), $\mathscr{N}_{\rm{w}}\big(\ket{\psi_{\rm{pas}}} \big)$ becomes predominantly dependent on $r$. 
On the other hand, in the case of SNS [Fig. \ref{fig_ncwe_pas_sns}(b)], we observe a monotonic dependence of $\mathscr{N}_{\rm{w}}$ on both $r$ and $m$. 
The apparent similarity between the curves of $\mathscr{N}_{\rm{w}}\big(\ket{\psi_{\rm{pas}}} \big)$ and $\mathscr{N}_{\rm{w}}\big(\ket{\psi_{\rm{sns}}} \big)$ for $m=1$ is due to the fact that, for $m=1$, both SNS and PAS are equivalent, i.e., $a^{\dagger}S(r)\ket 0=S(r)\ket 1$.
Hence, these states yield similar NC.
However for $\forall~ m\geq 2$, $\ket{\psi_{\rm{pas}}}$ and $\ket{\psi_{\rm{sns}}}$ are very different from each other, as discussed in \cite{nc_bsent_bose_kumar}.

\section*{VI. $\mathscr{N}_{\rm{w}}$ for Some Gaussian Mixed States: Squeezed Thermal States}

Any single mode Gaussian state, as elaborated in Eq. \ref{eq_def_smgs_dst}, could be written as displaced squeezed thermal state.
However, the NC of the state is independent of the global displacement, since, the Wehrl entropy remains invariant under any rigid translation, as discussed in Sec. 3.
Thus, while discussing NC of a Gaussian mixed state, it is sufficient to deal with squeezed thermal state only.
As described in Eq. (\ref{eq_def_wig_gauss}), any Gaussian state is well represented by its variance matrix $V$ satisfying the canonical relation (Eq. \ref{eq_vm_cond_bonafide}). 
A single mode squeezed thermal state is given by
\begin{equation}
\rho_{\rm{st}}= S_{\rm{sq}}(r) \rho_{\rm{th}}(\bar{n}) S^{\dagger}_{\rm{sq}}(r),
\label{eq_rho_st}
\end{equation}
where, $\bar{n}$ is the average number of photon in the thermal state and $r$ is the squeezed parameter.
For the sake of simplicity we consider real squeezing. 
The variance matrix of the squeezed thermal state is given by
\begin{equation}
V_{\rm{st}} = \begin{pmatrix}
\frac{e^{2r}\kappa}{2} & 0\\
0 & \frac{e^{-2r}\kappa}{2}
\end{pmatrix},
\label{eq_vm_st}
\end{equation}
where, $\kappa = 2\bar{n} + 1$.
For the variance matrix $V_{\rm{st}}$, its trace and determinant are given by
\begin{equation}
\epsilon_{\rm{st}} = \frac{\kappa}{2} \frac{e^{4r} + 1}{e^{2r}}~;~\delta_{\rm{st}} = \frac{\kappa^{2}}{4}.
\label{eq_epsilon_delta_st}
\end{equation}

Putting the expressions for $\epsilon_{\rm{st}}$ and $\delta_{\rm{st}}$ in Eq. (\ref{eq_meas_ncwe_gauss}), we get the NC of for $\rho_{\rm{st}}$ as
\begin{equation}
\mathscr{N}_{\rm{w}}(\rho_{\rm{st}}) = \max\big\lbrace 0, \frac{1}{2} \ln \big[\frac{(\kappa - 1)^{2} + 4 \kappa \mu^{2}}{2(\kappa^{2} + 1)} \big] \big\rbrace,
\label{eq_meas_ncwe_st}
\end{equation}
where, $\mu = \cosh r$.
Quite evidently, the condition of NC for Gaussian mixed state (Eq. \ref{eq_meas_ncwe_gauss}) leads to the condition $r>\frac{1}{2} \ln (2\bar{n}+1)$.

\begin{figure}[h!]
\includegraphics[scale=1.7]{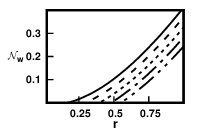}
\vspace*{-10pt}
\caption{Plot of $\mathscr{N}_{\rm{w}}$ vs $r$ for $\rho_{\rm{st}}$.
Different curves correspond to $\bar{n}=0.2$ (solid line), $0.4$ (dashed line), $0.6$ (dotted line), $0.8$ (dashed dotted line) and $1.0$ (dashed double dotted line).}
\label{fig_ncwe_sts}
\end{figure}

In Fig. \ref{fig_ncwe_sts}, we plot the NC of squeezed thermal state. 
Beyond the critical value of squeezed parameter $r$ ($\geq\frac{1}{2}\ln [2\bar{n}+1]$), $\mathscr{N}_{\rm{w}}(\rho_{\rm{st}})$ increases monotonically with increase in $r$. 
On the other hand, with increase in $\bar{n}$, $\mathscr{N}_{\rm{w}}(\rho_{\rm{st}})$ decreases.

In this context, it is worth looking at the corresponding pure state limit, i.e. the case of squeezed vacuum state that is obtained by considering $\bar{n}\rightarrow 0$ or $\kappa\rightarrow 1$ in Eq. (\ref{eq_meas_ncwe_st}). 
It is straightforward to check that
\begin{equation}
\lim_{\bar{n}\rightarrow 0} \mathscr{N}_{\rm{w}}(\rho_{\rm{st}}) = \ln \mu,
\end{equation}
the NC of squeezed vacuum state as obtained in Eq. (\ref{eq_meas_ncwe_sc}).

\section*{VII. Conclusion}

To summarize, in this paper, we have proposed a new quantification of the quantumness of single mode quantum optical states in terms of Wehrl entropy. 
We have quantified the quantumness of any state as the difference between its total Wehrl entropy and that arising due to its classical characteristics. 
We have obtained analytic expressions for a broad class of states, in particular, all pure states and Gaussian mixed states. 
Moreover, the proposed quantification is shown to be directly proportional to the NC of any Gaussian state, as inferred from its depth. 
The evaluation of NC for the non-Gaussian mixed states is subject to an extensive numerical computation that goes beyond the scope of the current paper. 
This, we shall address elsewhere.

In the case of nonclassical states of light, generated under single NC-inducing operation, $\mathscr{N}_{\rm{w}}$ quantifies the NC efficiently. 
It successfully distinguishes between the even and odd Schrodinger kittens (when coherent amplitude is small).
Besides, it shows that both the states are macroscopically equally nonclassical, irrespective of the parity, as observed in terms of the Wigner negativity \cite{ncm_nwf_kenfack}. 
Our quantification of NC also sheds light on the relative competition between the NC-inducing operations in the case of quantum optical states which are generated under multiple NC-inducing operations, as predicted in \cite{nc_bsent_bose_kumar}. 
In the case of mixed Gaussian nonclassical state, for example, a squeezed thermal state, it quantifies the NC of the state in line of the results obtained earlier with other well-known measures.

In recent times, Husimi-Kano $Q$ distribution, the classical like distribution, has gained much interest in both detection and quantification of the non-Gaussian character of any quantum optical state \cite{ngm_we_ivan, ngd_qdist_hughes}. 
%Miranowicz {\em et. al.} have further investigated Wehrl phase distribution in the context of distinguishing different pure state superpositions \cite{wepd_miranowicz}.
Similar approach using $SU(2)$ $Q$ distribution on the Poincare sphere has also been proposed to quantify the quantumness of a two-mode quantized electromagnetic field in terms of the polarization degrees \cite{su2_qdist_luis, su2_qdist_nc_luis}.
Here, we present a simple quantification of the quantumness for single mode quantum optical states in terms of a positive semi-definite quadrature distribution function, namely the Husimi $Q$ distribution, in contrast to the approaches based on phase-space singularity and/or negativity \cite{ncm_depth_lee, ncm_depth_crit_barnett, ncm_nwf_kenfack, ncm_kiesel, ncm_agudelo}.   
The efficacy of our proposal lies in the fact that in most of the cases it is could be computed analytically as well as the underlying distribution could be retrieved experimentally in optical heterodyne detection \cite{heterodyne_kimble}.

\section*{Acknowledgment}

Author is indebted to M. Sanjay Kumar and Samyadeb Bhattacharya in S. N. Bose National Centre for Basic Sciences, Kolkata, India for numerous discussion and critical remarks.
%Author also thanks anonymous referees for their insightful comment / suggestion (s) that has helped in formalizing the concept and improving the presentation of the manuscript.

\section*{Appendix: Wehrl Entropy of a Single Mode Gaussian State}

The Husimi-Kano $Q$ distribution could be written as a Gaussian convolution of the Wigner distribution as
\begin{equation}
Q(\beta,\beta^{*}) = 2 \int \frac{d^{2}\alpha}{\pi}~ W(\alpha,\alpha^{*})~ e^{-2|\beta - \alpha|^{2}},
\label{eq_def_WtoQ_complex}
\end{equation}
where, ($\alpha,\beta$) are the complex quadrature.
Writing, $\alpha = \frac{1}{\sqrt{2}} (R_{1} + iR_{2})$ and $\beta = \frac{1}{\sqrt{2}} (\tilde{R}_{1} + i\tilde{R}_{2})$, where, $R = (x,p)^{\rm{T}}$ and $\tilde{R} = (\tilde{x},\tilde{p})^{\rm{T}}$, we can recast Eq. (\ref{eq_def_WtoQ_complex}) in terms of the real quadrature as
\begin{equation}
Q(\tilde{R}) = 2 \int \frac{dR}{2\pi}~ W(R)~ e^{-(R-\tilde{R})^{\rm{T}}(R-\tilde{R})}.
\label{eq_def_WtoQ_real}
\end{equation}

Replacing $W(R)$ in Eq. (\ref{eq_def_WtoQ_real}) by the Gaussian Wigner distribution $W^{\rm{G}}(R)$ from Eq. (\ref{eq_def_wig_gauss}), we get
\begin{widetext}
\begin{align}
Q(\tilde{R})& = 2\sqrt{\det V^{-1}}~ \int\frac{dR}{2\pi}~ e^{-\frac{1}{2}(R-D)^{\rm{T}} V^{-1} (R-D)}~ e^{-(R-\tilde{R})^{\rm{T}}(R-\tilde{R})} \nonumber \\
& = 2\sqrt{\det V^{-1}}~ e^{\lbrace \tilde{R}^{\rm{T}}\tilde{R} + \frac{1}{2}\mathscr{D}^{\rm{T}} V^{-1} \mathscr{D} - \frac{1}{2}\mathscr{D}^{\rm{T}} M \mathscr{D}\rbrace}~ \int\frac{dR}{2\pi}~ e^{-\frac{1}{2}(R-M^{-1}\mathscr{D})^{\rm{T}}M(R-M^{-1}\mathscr{D})} \nonumber \\
& = 2\sqrt{\frac{\det V^{-1}}{\det M}}~ e^{\lbrace \tilde{R}^{\rm{T}}\tilde{R} + \frac{1}{2}\mathscr{D}^{\rm{T}} V^{-1} \mathscr{D} - \frac{1}{2}\mathscr{D}^{\rm{T}} M \mathscr{D}\rbrace},
\end{align}
\end{widetext}
where, $M = V^{-1} + 2I$ and $\mathscr{D} = 2\tilde{R} + V^{-1}D$.
Let's consider a $2\times 2$ real symmetric matrix quantum variance matrix $V$ of the form
\begin{equation}
V = \begin{pmatrix}
v_{11} & v_{12}\\
v_{12} & v_{22}
\end{pmatrix}
\end{equation}
that satisfies the condition as described in Eq. (\ref{eq_vm_cond_bonafide}). 
A straightforward calculation yields
\begin{equation}
Q(\tilde{R}) = \sqrt{\det \mathscr{M}}~ e^{\frac{1}{2}(\tilde{R}-D)^{\rm{T}}\mathscr{M}(\tilde{R}-D)},
\label{eq_def_qdist_gauss}
\end{equation}
where,
\begin{align}
\mathscr{M}& = 2(I - 2M^{-1}) \nonumber \\
& = \frac{2}{4\delta + 2\epsilon + 1}\begin{pmatrix}
2v_{22} + 1 & -2v_{12} \\
-2v_{12} & 2v_{11} + 1
\end{pmatrix}
\end{align}

The quantities $\delta$ and $\epsilon$ are defined to be $\delta = \det V = v_{11}v_{22} - v_{12}^{2}$ and $\epsilon = \rm{Tr}[V] = v_{11} + v_{22}$.
Quite evidently, the Wehrl entropy of the single mode Gaussian state $\rho^{\rm{G}}$, described by the variance matrix $V$, is given by
\begin{align}
H_{\rm{w}}(\rho^{\rm{G}})& = -\int\frac{d\tilde{R}}{2\pi}~Q^{\rm{G}}(\tilde{R})~ \ln Q^{\rm{G}}(\tilde{R}) \nonumber \\
& =  1 - \ln \sqrt{\det \mathscr{M}} \nonumber \\
& =  1 + \frac{1}{2} \ln \big[\frac{4\delta + 2\epsilon + 1}{2} \big].
\end{align}

\end{document}